\documentclass{article}
\usepackage[utf8]{inputenc}
\usepackage{amsfonts}
\usepackage{amsmath}

\newcommand{\E}{\mathbb{E}}
\newcommand{\norm}[1]{\left\lVert#1\right\rVert}

\newcommand{\eqdef}{\mathrel{\stackrel{\makebox[0pt]{\mbox{\normalfont\tiny def}}}{=}}}

\newcommand\opn{\mathrel{\ooalign{$\subseteq$\cr
  \hidewidth\raise.225ex\hbox{$\circ\mkern.5mu$}\cr}}}

\DeclareMathOperator*{\argmin}{arg\,min}
\usepackage{xcolor}

\usepackage{tikz}
\usepackage{caption}
\usepackage{subcaption}
\usepackage{tabu}
\PassOptionsToPackage{numbers}{natbib}

\usepackage[final]{neuripsDeepInverse_2020}

\usepackage[utf8]{inputenc} %
\usepackage[T1]{fontenc}    %
\usepackage{hyperref}       %
\usepackage{url}            %
\usepackage{booktabs}       %
\usepackage{amsfonts}       %
\usepackage{nicefrac}       %
\usepackage{microtype}      %

\title{Generative Tomography Reconstruction}

\author{
  Matteo Ronchetti\\
  University of Pisa \\
  \texttt{mttronchetti@gmail.com} \\
  \And
  Davide Bacciu \\
  University of Pisa \\
  \texttt{bacciu@di.unipi.it} \\
}

\begin{document}

\maketitle

\begin{abstract}
  We propose an end-to-end differentiable architecture for tomography reconstruction that directly maps a noisy sinogram into a denoised reconstruction. Compared to existing approaches our end-to-end architecture produces more accurate reconstructions while using less parameters and time.  We also propose a generative model that, given a noisy sinogram, can sample realistic reconstructions. This generative model can be used as prior inside an iterative process that, by taking into consideration the physical model, can reduce artifacts and errors in the reconstructions.  
\end{abstract}

\section{Introduction}
This work tackles a prototypical linear inverse problem: reconstruction of Computed Tomography (CT) images in the presence of large amount of noise. The goal of CT reconstruction is to infer an image $x$, given a noisy sinogram $y_\delta \sim n(Ax)$, where $A$ is the discrete Radon transform \cite{radon_discrete} and $n(\cdot)$ is the noise distribution.
Solving such a reconstruction problem is difficult due to the ill-conditioning of Radon transforms. It is also a problem of high practical relevance and impact, given the widespread use of CT algorithms in applications such as medical imaging \cite{dental_ct, breast_ct, electron_ct}, luggage scanning \cite{luggage_ct} and damage detection in concrete structures \cite{concrete_ct}.

This work is centred around two original contributions. First, we design an end-to-end differentiable architecture that directly maps a noisy sinogram to a denoised reconstruction. By combining reconstruction and denoising functionalities into a single model, we are able to improve reconstruction quality while reducing inference time. Existing ML approaches to noisy CT reconstruction \cite{cnn_ct, cnn_ct:2, cnn_ct:3} use a convolutional neural network (CNN) to denoise a reconstruction computed by a non-adaptive algorithm. Contrary to that, we apply the learned reconstruction and denoising process directly to the sinogram. This allows the model to reduce noise before it is amplified by the reconstruction process. The second contribution leverages the differentiable architecture to propose a generative model that can sample reconstructions which are realistic and compatible with sensor readings. The reconstructions produced by the proposed generative model can be iteratively improved by taking into consideration the physical model of sinogram generation. This iterative algorithm can reduce reconstruction errors and biases, guaranteeing that the produced solution is compatible with what is measured by the sensors. GAN-based approaches in literature for solving linear inverse problems \cite{gan_iter, gan_iter:2, gan_iter:3} compute reconstructions iteratively by improving an initial random guess. In contrast to that, our approach can generate good solutions even without iterating, making the iterative improvement process much faster. As a final note, we point out the generality of our approaches, which allow a straightforward application to other linear inverse problems, such as limited/sparse angle tomography, image super-resolution and image deblurring. 

\section{Sinogram-Based Tomography Reconstruction}
Before delving into the details of the two contributions, we briefly summarize the noise model used to simulate sensor readings. We take into consideration shot and electronic noise, and quantization errors as follows
\begin{align*}
    z \sim \text{Pois}(\exp(s - y)) + N(0, \epsilon) \\
    r = \text{clamp}(\text{round}(z/k), 0, 2^b-1)
\end{align*}
where $\exp(s)$ is the X-ray intensity, $b$ is the number of bits used by the detector and $k$ is a scaling parameter. We assume that, given readings $r$ (that are integer values), sinograms are distributed as $y|r \sim \mathcal{N}(\mu(r), \text{diag}(\sigma^2(r)))$, where $\mu(\cdot)$ and $\sigma(\cdot)$ can be modelled by CNNs. We train $\mu(\cdot)$ and $\sigma(\cdot)$ networks to minimize the negative log-likelihood (refer to Appendix \ref{app:noise_model} for more details).

\subsection{Reconstruction Model}
\label{e2e}
Inspired by filtered back-projection (FBP) \cite{fbp_filtering}, which reconstructs $x$ as $\widetilde{x} = A^T f(y)$, where $f(\cdot)$ is a fixed linear filtration, we design our reconstruction model as $g(r) = g_2(A^T g_1(r))$ where both $g_1$ and $g_2$ are trainable neural networks. 
The Radon transform is ill-conditioned and FBP is well-known to greatly amplify noise. Our rationale is that by combining reconstruction and denoising, we allow the model to mitigate the reconstruction errors caused by the ill-conditioning of the Radon transform. Furthermore, since the output of $g_1(r)$ is not limited to be a single channel, it allows richer information flow from the sinogram to the image.

Briefly, our reconstruction model is composed of two U-nets \cite{unet} connected by Radon backprojection (multiplication by $A^T$). The first one ($g_1$) processes the sensor readings and outputs a multi-channel sinogram, each channel is transformed into an image using Radon backprojection and fed to the second network. 
The first layer of $g_2$ concatenates to its input (of shape $(\text{ch}, 256, 256)$) a 4 dimensional positional embedding (of shape $(4, 256, 256)$) which is used by the network to encode pixel coordinates. 
The down-sampling blocks of the U-net are composed of a strided convolution followed by a sequence of residual blocks \cite{resnet}. Each U-net up-sampling block takes the output of the previous block, double its resolution using bilinear up-sampling, and concatenates it with the skip connection coming from the down-sampling block at the corresponding resolution. Following, the concatenation channels are weighted using channel attention \cite{channel_attention} and inputted to a $1\times 1$ convolution that shrinks their number. Finally they are processed by a sequence of residual blocks. A detailed account of the architecture is provided in Appendix \ref{app:model_arch}.

The key point in our convolutional scheme is that the Radon backprojection between the two U-nets is differentiable. Therefore the model discussed above can be trained end-to-end by jointly optimizing $g_1$ and $g_2$ to minimize an image reconstructions loss.

\subsection{Generative Model}
\label{sec:gan}
We design an end-to-end generator model that, given noisy sensor readings $r$, can sample different possible reconstructions. We show how, by combining it with a discriminator and the noise model, it can be used to iteratively improve reconstructions. 

We build on the Wasserstein GAN (WGAN) \cite{wgan} framework and train a generator $G$ to minimize
\begin{equation}
    \label{generator_obj}
\E_{r, z}\left[\norm{AG(r, z) - \mu(r)}_{\sigma(r)}^2 + \lambda D(G(r, z))\right] \text{ where } \norm{v}_{\sigma(r)}^2 \eqdef \sum_i \left(\frac{v_i}{\sigma_i(r)}\right)^2
\end{equation}
where $z$ is uniformly distributed over the sphere and $\lambda$ is a fixed regularization parameter.
Simultaneously, we train a discriminator $D$ to minimize
$$
\E_x[D(x)] - \E_{r,z}[D(G(r, z))] \quad\text{subject to}\quad \norm{D}_L \leq c
$$
where the Lipschitz-norm constraint on $D$ is enforced using Spectral Normalization \cite{spectral_normalization}. 

The generator $G(r, z)$ is an end-to-end model following the architecture described in Section \ref{e2e}. To make the computation of $\nabla_z G(r, z)$ efficient, we feed $z$ to the third last block of the $g_2$ network, so that back-propagating the gradient to $z$ only requires to go through few layers. The structure of this modified U-net block is depicted in Figure \ref{fig:zblock} to highlight the additional input from $z$.
The first layer of this block concatenates its inputs (by applying bilinear up-sampling to the ones coming from the previous block) and applies a $3\times3$ convolution followed by a PReLu activation \cite{prelu}.
The effect of $z$ on the output should be modulated depending on the amount of noise and uncertainty in the reconstruction. Therefore the last $32$ channels produced by the first layer of the block are fed to an exponential activation and used to scale the elements of $z$ (bottom part in Figure \ref{fig:zblock}). The remaining channels are then concatenated with the scaled version of $z$ and fed to a sequence of residual blocks. The parameter $z$ has shape $(32, 16, 16)$ and it is up-scaled to $(32, 64, 64)$ using bilinear interpolation.

If we train the generator using the loss described in (\ref{generator_obj}) it learns to ignore $z$. To avoid this problem, we regularize the output of $G$ by pushing the distribution of $A G(r, z)$ towards $\mathcal{N}(\mu(r), \text{diag}(\sigma^2(r)))$. During training we generate two different reconstructions for each noisy reading. Then, we fit a normal distribution with diagonal covariance on the two corresponding sinograms and compare it against $\mathcal{N}(\mu(r), \text{diag}(\sigma^2(r)))$ using the Kullback–Leibler divergence
\begin{equation*}
\begin{gathered}
    y_1 = AG(r, z_1) \qquad y_2 = AG(r, z_2) \\
    \norm{\frac{y_1 + y_2}{2} - \mu(r)}_{\sigma(r)}^2 + \sum_i \left[ 2\log\left(\frac{\sigma(r)}{\sigma_p}\right) + \left(\frac{\sigma_p}{\sigma(r)}\right)^2 \right] \text{ where } \sigma^2_p = \frac{(y_1 - y_2)^2}{2} \,.
\end{gathered}
\end{equation*}
Using this loss in place of the first term of (\ref{generator_obj}) solves the mode collapse problem without reducing training stability.

As a final note, the reconstruction produced by our generator can be iteratively improved using projected gradient to minimize $\norm{AG(r, z) - \mu(r)}_{\sigma(r)}^2 + \lambda D(G(r, z))$ subject to $\norm{z} = 1$. 
\begin{figure}
\begin{minipage}{.48\textwidth}
\centering
\tikzset{
  block/.style    = {draw, thick, rectangle, minimum height = 0.75cm,
    minimum width = 0.75cm},
  var/.style      = {draw, circle, node distance = 1cm}
}
 \hspace*{-0.5cm}
\begin{tikzpicture}[auto, thick, node distance=2.0cm, inner sep=0.1cm]
\draw
  node at (-0.25,0) [] (in) {}
	node at (0.65,0) [block] (conv) {Conv}
  node at (1.85,0)  [block] (split) {Split}
  node at (1.85,-1) [block] (exp) {exp}
  node at (3.2, -1) [block] (mul) {\Large$\times$}
  node at (3.2,-2) [block] (upscale) {Upscale}
  node at (3.2,-2.85) [var] (z) {\large$z$}
  node at (3.2,0) [block] (cat) {Concat}
  node at (5.0,0) [block] (res) {Res Blocks}
  node at (6.4, 0) [] (out) {};
  \draw[->] (in) to (conv);
  \draw[->] (conv) to (split);
  \draw[->] (split) to (cat);
  \draw[->] (split) to (exp);
  \draw[->] (exp) to (mul);
  \draw[->] (z) to (upscale);
  \draw[->] (upscale) to (mul);
  \draw[->] (mul) to (cat);
  \draw[->] (cat) to (res);
  \draw[->] (res) to (out);
\end{tikzpicture}
  \caption{Structure of the modified U-net block used to insert $z$ into the generator.}
  \label{fig:zblock}
\end{minipage}%
\hfill
\begin{minipage}{.48\textwidth}
{
\tabulinesep=1mm
\begin{tabu}{ |c|c|c|c|c| } 
 \hline
 $g_1$ & $g_2$ & SSIM & Parameters & FPS \\ 
 \hline
 FBP & L & 76.6\% & $6.30 \cdot 10^6$ & 211 \\ 
 FBP & XL & 76.7\% & $9.37 \cdot 10^6$ & 178 \\
 FBP & XXL & 76.9\% & $1.56 \cdot 10^7$ & 121 \\
 \hline
 XXS & S-64 & 76.7\% & $2.63 \cdot 10^6$ & 239 \\ 
 XS & M & 76.9\% & $4.12 \cdot 10^6$ & 187 \\ 
 S & L & 77.4\% & $6.93 \cdot 10^6$ & 152 \\ 
 \hline
\end{tabu}
}
\caption{Comparison of denoising results using FBP based models (top) and end-to-end models (bottom).}
  \label{fig:results_table}
\end{minipage}
\end{figure}

\section{Experimental Assessment}
We use images from the DeepLesion dataset \footnote{\url{https://nihcc.app.box.com/v/DeepLesion}}, cropped to guarantee that pixels outside of the inscribed circle are zero and rescaled to $256\times 256$ (see Appendix \ref{app:dataset} for more details about preprocessing). We split train/validation/test by patient identifier and use multiple signal intensities (therefore noise levels) during training and testing. No augmentation is used, sensor readings are not fixed but randomly simulated before each training iteration.

We compare the reconstruction quality obtained by our end-to-end approach and an FBP model, also varying the dimensions of the U-net components $g_1$ and $g_2$. We use conventional names (XXS, XS, S, S-64, M, L, XL, XXL) to refer to increasing U-net dimensions (refer to Appendix \ref{app:model_arch} for further details). All the models are trained for $10$ epochs using the RAdam \cite{radam} optimizer, batch size of $16$ and mixed precision training. The learning rate is exponentially increased from $0$ to $3 \cdot 10^{-4}$ during the first $5000$ batches, then halved every $80000$ batches.

We measure the quality of reconstructions using structured similarity (SSIM) and compute inference speed as the number of frames per second (FPS) obtained with the largest possible batch size on a Tesla V100 GPU.
Results are shown in Table \ref{fig:results_table}, from which it can be noticed that end-to-end models achieve a more accurate reconstruction than FBP based models at comparable number of FPS. End-to-end models scale better, a $+0.7\%$ improvement in SSIM requires to increase the number of parameters $2.6\times$, compared to FBP models where an improvement of $+0.3\%$ costs $2.5\times$ more parameters. Furthermore end-to-end models are much more parameter-efficient, given that their largest model is only slightly larger than the smallest FBP model.
\begin{figure}
\centering
\includegraphics[width=0.49\textwidth]{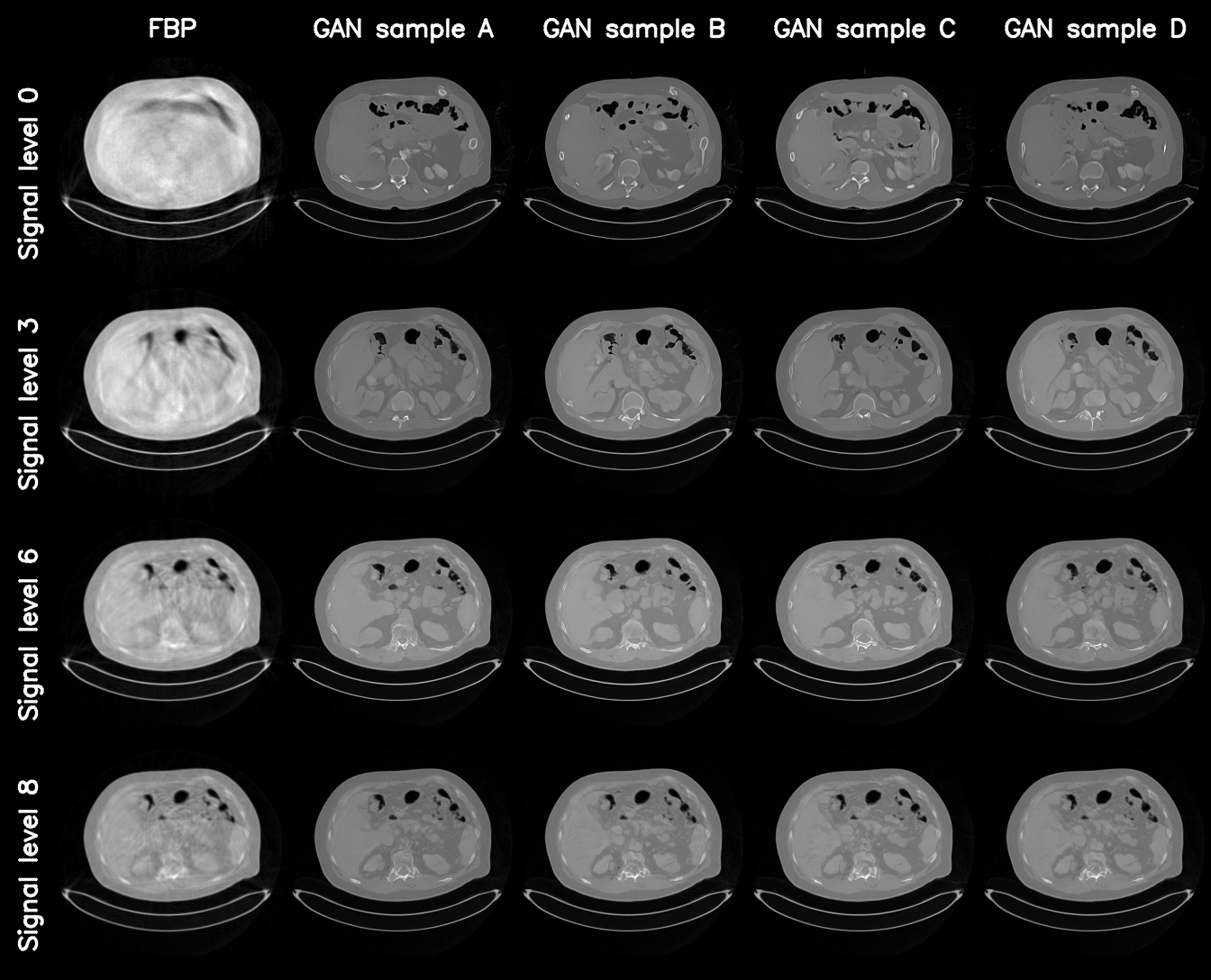}
\hfill
\includegraphics[width=0.49\textwidth]{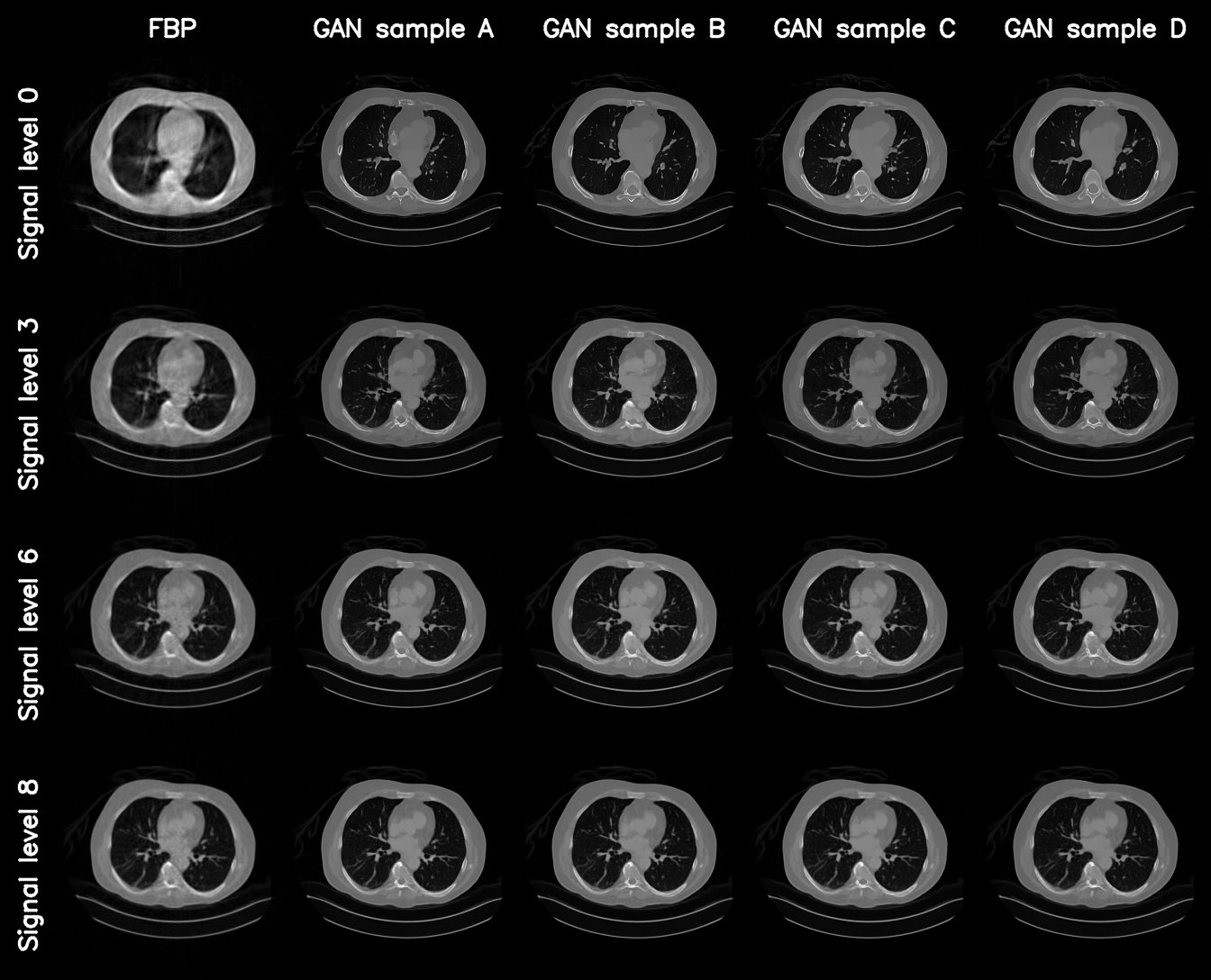}
\caption{Reconstruction results using FBP on the estimated sinogram $\mu(r)$ compared with 4 samples generated by the proposed GAN. Notice that, as signal increases, GAN reconstructions become more similar to one another.}
\label{fig:gan_samples}
\end{figure}

\begin{figure}
    \centering
    \includegraphics[width=0.7\textwidth]{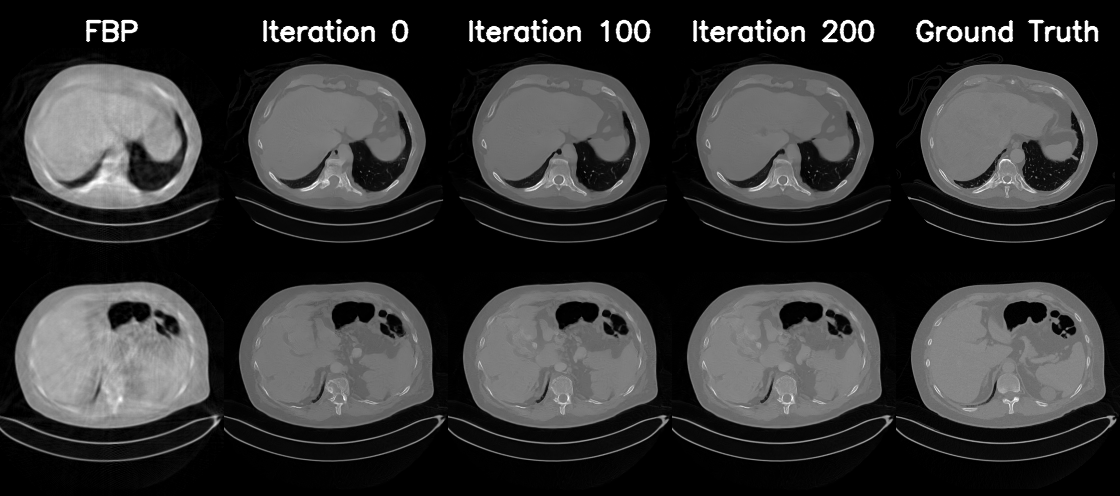}
    \caption{Progress of our iterative algorithm on two candidate reconstructions}
    \label{fig:gan_iter}
\end{figure}
Lower signal intensities (therefore higher noise) are used to simulate readings for the training of our generative model. Figure \ref{fig:gan_samples} depicts reconstructions computed by FBP and sampled from our GAN. Notice that GAN reconstructions looks realistic even with very high noise values and, as the noise decreases, converge to the correct reconstruction.

Figure \ref{fig:gan_iter} depicts the progress of the projected gradient algorithm on two candidate reconstructions using a learning rate of $10^{-4}$. Notice in particular how the reconstruction of the spinal cord is improved by the iterative algorithm.

\section{Conclusions and Future Work}
We proposed an end-to-end architecture for tomography reconstruction characterized by higher quality and performance than FBP-based denoising models.
The key point of our approach is the incorporation of the Radon transform as a differentiable component of the neural architecture, allowing to build the reconstruction directly from the sinogram. The end-to-end architecture can be used to design generative models capable of sampling multiple possible reconstructions. Also, we show that by taking into consideration the physical model, it is possible to iteratively improve the reconstructions computed by the generative model.
Our work can be easily extended to sparse angle and limited angle tomography or to other linear inverse problems.

\bibliographystyle{plain}
\bibliography{bibliography}

\appendix
\section{Noise Model}
\label{app:noise_model}

Without taking into consideration noise the ideal sensor readings would be $\exp(s - y)$, where $\exp(s)$ is the intensity of the X-ray emitted by the machine. In practice due to shot noise and electronic noise readings can be modelled as
$$
z \sim \text{Pois}(\exp(s - y)) + \mathcal{N}(0, \epsilon) \,.
$$
In this setting noise depends on the signal intensity (the stronger the signal the weaker the noise) and on the variance of the normal distribution that emulates electronic noise. Notice that, because of the Poisson distribution, noise will not be uniform across the sinogram and will be higher where $y$ is large.

In practice the analog value at each pixel of the sensor is converted into a digital integer value, this process creates quantization error. If readings are stored as $b$ bits values then this conversion process can be modelled as
$$
r = \text{clamp}(\text{round}(z/k), 0, 2^b-1)
$$
where $k$ is a normalization value that depends on the scale of the measured signal.

Given sensor readings $r$ (a matrix of integers in range $[0, 2^b)$) we would like to estimate the posterior probability over noiseless sinograms $\pi(y | r)$. We make the simplifying assumption of $y|r \sim \mathcal{N}(\mu(r), \text{diag}(\sigma^2(r)))$, where $\mu(\cdot)$ and $\sigma(\cdot)$ are CNNs. Because both the mean and variance are modelled by CNNs intra-pixel relationships can be modelled.

We first train $\mu(\cdot)$ to minimize
$$
\E[\norm{y - \mu(r)}_2^2]
$$
and, after $\mu(\cdot)$ is trained, we use it to train  $\sigma(\cdot)$ to minimize the negative log likelihood
$$
\E \left[ \frac{(y - \mu(r))^2}{\sigma^2(r)} + 2\log(\sigma(r))  \right]\, .
$$
Finally we use knowledge distillation to combine $\mu(\cdot)$ and $\sigma(\cdot)$ into a single CNN.

\section{Model Architecture}
\label{app:model_arch}
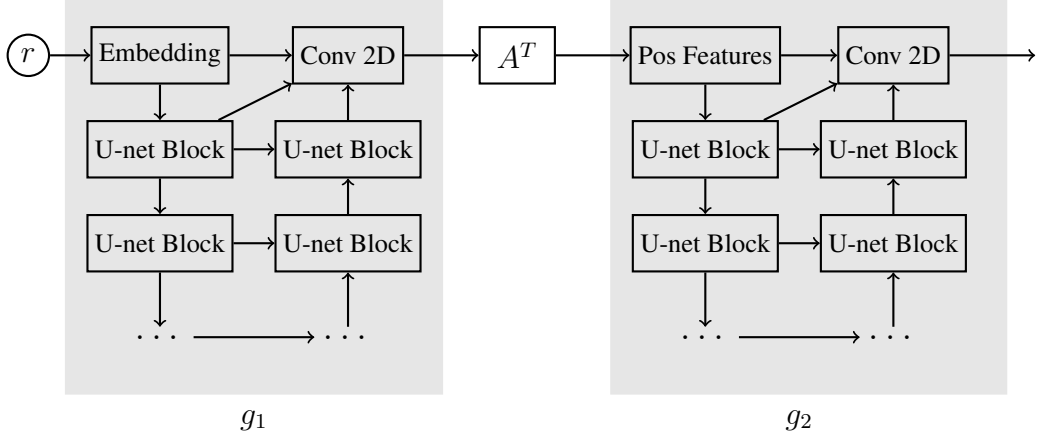
\begin{figure}
  \centering
  \tikzset{
  block/.style    = {draw, thick, rectangle, minimum height = 0.75cm,
    minimum width = 1cm},
  var/.style      = {draw, circle, node distance = 1cm}
}
\definecolor{offwhite}{gray}{0.90}
\begin{tikzpicture}[auto, thick, node distance=2.0cm, inner sep=0.1cm]
  \draw [color=offwhite, fill](0.25,0.75) rectangle (5.25,-4.5);
 \node at (2.75,-4.85) [] {\large$g_1$};
 \draw [color=offwhite, fill](7.5,0.75) rectangle (12.75,-4.5);
 \node at (10,-4.85) [] {\large$g_2$};
\draw
  node at (-0.25,0) [var] (r) {\large$r$}
  node at (1.5,0) [block] (emb) {Embedding}
  node at (1.5,-1.25) [block] (s_x1) {U-net Block}
  node at (1.5,-2.5) [block] (s_x2) {U-net Block}
  node at (1.5,-3.75) [] (s_x3) {\Large$\dots$}
  node at (4,-3.75) [] (s_u3) {\Large$\dots$}
  node at (4,-2.5) [block] (s_u2) {U-net Block}
  node at (4,-1.25) [block] (s_u1) {U-net Block}
  node at (4,0) [block] (s_readout) {Conv 2D}
  node at (6.25,0) [block] (radon) {\large$A^T$}
  node at (8.75,0) [block] (pos) {Pos Features}
  node at (8.75,-1.25) [block] (x1) {U-net Block}
  node at (8.75,-2.5) [block] (x2) {U-net Block}
  node at (8.75,-3.75) [] (x3) {\Large$\dots$}
  node at (11.25,-3.75) [] (u3) {\Large$\dots$}
  node at (11.25,-2.5) [block] (u2) {U-net Block}
  node at (11.25,-1.25) [block] (u1) {U-net Block}
  node at (11.25,0) [block] (readout) {Conv 2D}
  node at (13.25,0) [] (out) {};
  \draw[->] (r) to (emb);
  \draw[->] (emb) to (s_x1);
  \draw[->] (s_x1) to (s_x2);
  \draw[->] (s_x2) to (s_x3);
  \draw[->] (s_x3) to (s_u3);
  \draw[->] (s_u3) to (s_u2);
  \draw[->] (s_x2) to (s_u2);
  \draw[->] (s_u2) to (s_u1);
  \draw[->] (s_x1) to (s_u1);
  \draw[->] (emb) to (s_readout);
  \draw[->] (s_x1) to (s_readout);
  \draw[->] (s_u1) to (s_readout);
  \draw[->] (s_readout) to (radon);
  \draw[->] (radon) to (pos);
  \draw[->] (pos) to (x1);
  \draw[->] (x1) to (x2);
  \draw[->] (x2) to (x3);
  \draw[->] (x3) to (u3);
  \draw[->] (u3) to (u2);
  \draw[->] (x2) to (u2);
  \draw[->] (u2) to (u1);
  \draw[->] (x1) to (u1);
  \draw[->] (pos) to (readout);
  \draw[->] (x1) to (readout);
  \draw[->] (u1) to (readout);
  \draw[->] (readout) to (out);
\end{tikzpicture}
  \caption{Architecture of the end-to-end model used for simultaneous reconstruction and denoising.}
  \label{fig:e2e_detail}
\end{figure}

\begin{figure}
  \centering
  \tikzset{
  block/.style    = {draw, thick, rectangle, minimum height = 0.75cm,
    minimum width = 1cm},
  var/.style      = {draw, circle, node distance = 1cm}
}
\definecolor{offwhite}{gray}{0.90}
\begin{tikzpicture}[auto, thick, node distance=2.0cm, inner sep=0.1cm]
  \draw [color=offwhite, fill](0.25,0.75) rectangle (5.25,-4.5);
 \node at (2.75,-4.85) [] {\large$g_1$};
 \draw [color=offwhite, fill](7.5,0.75) rectangle (12.75,-4.5);
 \node at (10,-4.85) [] {\large$g_2$};
\draw
  node at (-0.25,0) [var] (r) {\large$r$}
  node at (1.5,0) [block] (emb) {Embedding}
  node at (1.5,-1.25) [block] (s_x1) {U-net Block}
  node at (1.5,-2.5) [block] (s_x2) {U-net Block}
  node at (1.5,-3.75) [] (s_x3) {\Large$\dots$}
  node at (4,-3.75) [] (s_u3) {\Large$\dots$}
  node at (4,-2.5) [block] (s_u2) {U-net Block}
  node at (4,-1.25) [block] (s_u1) {U-net Block}
  node at (4,0) [block] (s_readout) {Conv 2D}
  node at (6.25,0) [block] (radon) {\large$A^T$}
  node at (8.75,0) [block] (pos) {Pos Features}
  node at (8.75,-1.25) [block] (x1) {U-net Block}
  node at (8.75,-2.5) [block] (x2) {U-net Block}
  node at (8.75,-3.75) [] (x3) {\Large$\dots$}
  node at (11.25,-3.75) [] (u3) {\Large$\dots$}
  node at (11.25,-2.5) [block] (u2) {Modified Block}
  node at (13.25,-2.5) [var] (z) {\Large$z$}
  node at (11.25,-1.25) [block] (u1) {U-net Block}
  node at (11.25,0) [block] (readout) {Conv 2D}
  node at (13.25,0) [] (out) {};
  \draw[->] (r) to (emb);
  \draw[->] (emb) to (s_x1);
  \draw[->] (s_x1) to (s_x2);
  \draw[->] (s_x2) to (s_x3);
  \draw[->] (s_x3) to (s_u3);
  \draw[->] (s_u3) to (s_u2);
  \draw[->] (s_x2) to (s_u2);
  \draw[->] (s_u2) to (s_u1);
  \draw[->] (s_x1) to (s_u1);
  \draw[->] (emb) to (s_readout);
  \draw[->] (s_x1) to (s_readout);
  \draw[->] (s_u1) to (s_readout);
  \draw[->] (s_readout) to (radon);
  \draw[->] (radon) to (pos);
  \draw[->] (pos) to (x1);
  \draw[->] (x1) to (x2);
  \draw[->] (x2) to (x3);
  \draw[->] (x3) to (u3);
  \draw[->] (u3) to (u2);
  \draw[->] (x2) to (u2);
  \draw[->] (z) to (u2);
  \draw[->] (u2) to (u1);
  \draw[->] (x1) to (u1);
  \draw[->] (pos) to (readout);
  \draw[->] (x1) to (readout);
  \draw[->] (u1) to (readout);
  \draw[->] (readout) to (out);
\end{tikzpicture}
  \caption{Architecture of the generator.}
  \label{fig:generator}
\end{figure}
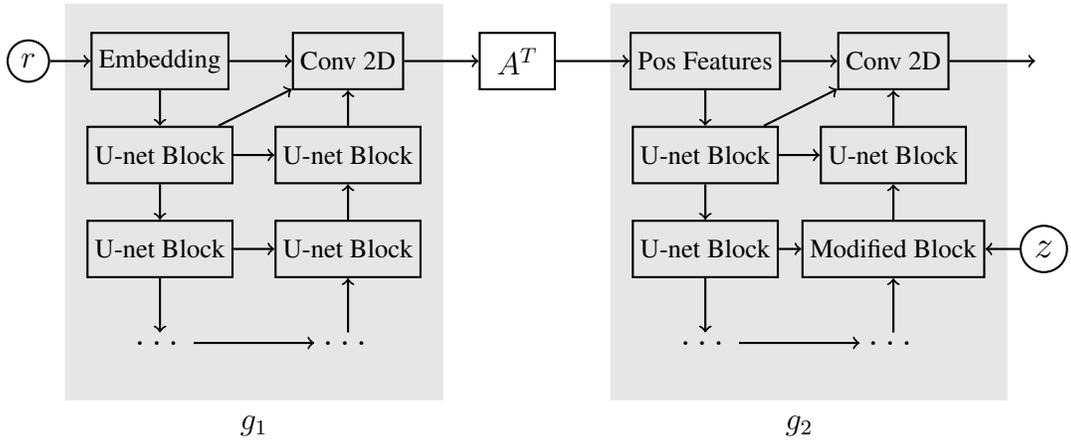

The architecture of the end-to-end reconstruction model is depicted in Figure \ref{fig:e2e_detail}. Both $g_1$ and $g_2$ are instances of the U-net architecture \cite{unet}.
Blocks that have more than one input, first up-sample each input to the largest resolution using bilinear up-sampling, then concatenate all the inputs on the channel dimension. \\
The architecture of the proposed generator is shown in Figure \ref{fig:generator} and is analogous to the end-to-end reconstruction model. The only difference is the modified block that is used to receive $z$ (bottom right of Figure \ref{fig:generator}) which is described in Section \ref{sec:gan}.

The down-sampling blocks of the U-net are composed of a strided convolution followed by a sequence of residual blocks \cite{resnet}. Each U-net up-sampling block takes the output of the previous block, double its resolution using bilinear up-sampling, and concatenates it with the skip connection coming from the down-sampling block at the corresponding resolution. Following, the concatenation channels are weighted using channel attention \cite{channel_attention} and inputted to a $1\times 1$ convolution that shrinks their number. Finally they are processed by a sequence of residual blocks.

We use conventional names (XXS, XS, S, S-64, M, L, XL, XXL) to refer to increasing U-net dimensions, the only difference between S and S-64 is the number of channels.
The number of channels used by the residual blocks is doubled every two strided convolutions. The first block of XXS, XS, S uses 32 channels, while the first block of S-64, M, L, XL, XXL uses 64 channels. The number of residual blocks used inside each U-net block is shown in the following table:
\begin{table}
\centering
\tabulinesep=1.5mm
\begin{tabu}{|l|ccccccc|}
\hline
Resolution & XXS & XS & S & M & L & XL & XXL \\
\hline
256 & 1 & 1 & 1 & 2 & 2 & 3 & 5 \\
128 & 1 & 1 & 1 & 1 & 2 & 3 & 5 \\
64 & 2 & 1 & 1 & 1 & 2 & 3 & 5 \\
32 & - & 2 & 2 & 1 & 2 & 3 & 5 \\
16 & - & - & 2 & 1 & 2 & 3 & 5 \\
8 & - & - & - & 2 & 3 & 5 & 9 \\
16 & - & - & - & 2 & 3 & 4 & 6 \\
32 & - & - & 2 & 2 & 3 & 4 & 6 \\
64 & - & 2 & 2 & 2 & 3 & 4 & 6 \\
128 & 2 & 2 & 2 & 2 & 3 & 4 & 6 \\
256 & 1 & 2 & 2 & 2 & 4 & 4 & 6 \\
\hline
\end{tabu}
\end{table}

In end-to-end models the latest convolution of $g_1$ reduces the number of channels before Radon backprojection. The smallest end-to-end model (which uses a XXS U-net for $g_1$ and a S-64 U-net for $g_2$) uses 16 channels for Radon backprojection, while the other two end-to-end models use 32 channels.

\section{Motivation for the Generator Loss}

A standard approach for solving inverse problems is to use regularization, this amounts to solving
$$
\argmin_x \norm{Ax - y}_2^2 + \lambda D(x)
$$
where $D(x)$ is a penalty term that promotes desired properties of $x$ and $\lambda$ is the regularization parameter.

First, we modify this formulation to take into consideration our sinogram estimation:
$$
\argmin_x \norm{Ax - \mu(r)}_{\sigma(r)}^2 + \lambda D(x)
$$
where
$$
\norm{Ax - \mu(r)}_{\sigma(r)}^2 \eqdef \sum \frac{(y - \mu(r))^2}{\sigma^2(r)}
$$
With this formulation we are taking into consideration the non-uniform confidence in the estimated sinogram values. There is still the problem that we are directly optimizing over images, and we still have to define $D(x)$.

Instead of optimizing $x$ we introduce a generator $G(r, z)$ and optimize over the free parameter $z$ that lies on the $m+1$ dimensional sphere:
$$
\argmin_{z \in S^m} \norm{AG(r, z) - \mu(r)}_{\sigma(r)}^2 + \lambda D(G(r, z)) \,.
$$

\section{Dataset Preprocessing}
\label{app:dataset}
We use the DeepLesion dataset (https://nihcc.app.box.com/v/DeepLesion) which is a large dataset of annotated tomographic images stored with 16 bit precision. We only use the images and don't consider the annotations for training our model. Images are preprocessed as follows:
\begin{enumerate}
  \item Subtract 32768 from the pixel intensity to obtain the original Hounsfield unit (HU) values. HU is a measure of radiodensity where air has density $-1000$HU and water $0$HU.
  \item Create a binary version of the image by thresholding pixel with value larger than $-9050$HU (threshold is a bit higher than the density of air).
  \item Find the smallest circle that contains all the pixels of the binarized image and to crop the corresponding square region from the image.
  \item If the cropped region is smaller than $256\times256$ discard it, otherwise rescale to $256\times256$ and add to the dataset.
\end{enumerate}

\section{Implementation Details}
Training the model requires computation and differentiation of Radon forward and backward projections (corresponding respectively to products with $A$ and $A^T$). These operations are linear and therefore differentiable, we use the implementation offered by the TorchRadon library \cite{torch_radon} which is integrated with PyTorch \cite{pytorch} and allows the backpropagation of gradients.

\section{Additional Generated Reconstructions}
The following figures contain additional images generated by our model using all the tested signal levels.

\begin{figure}
    \centering
    \includegraphics[width=\textwidth]{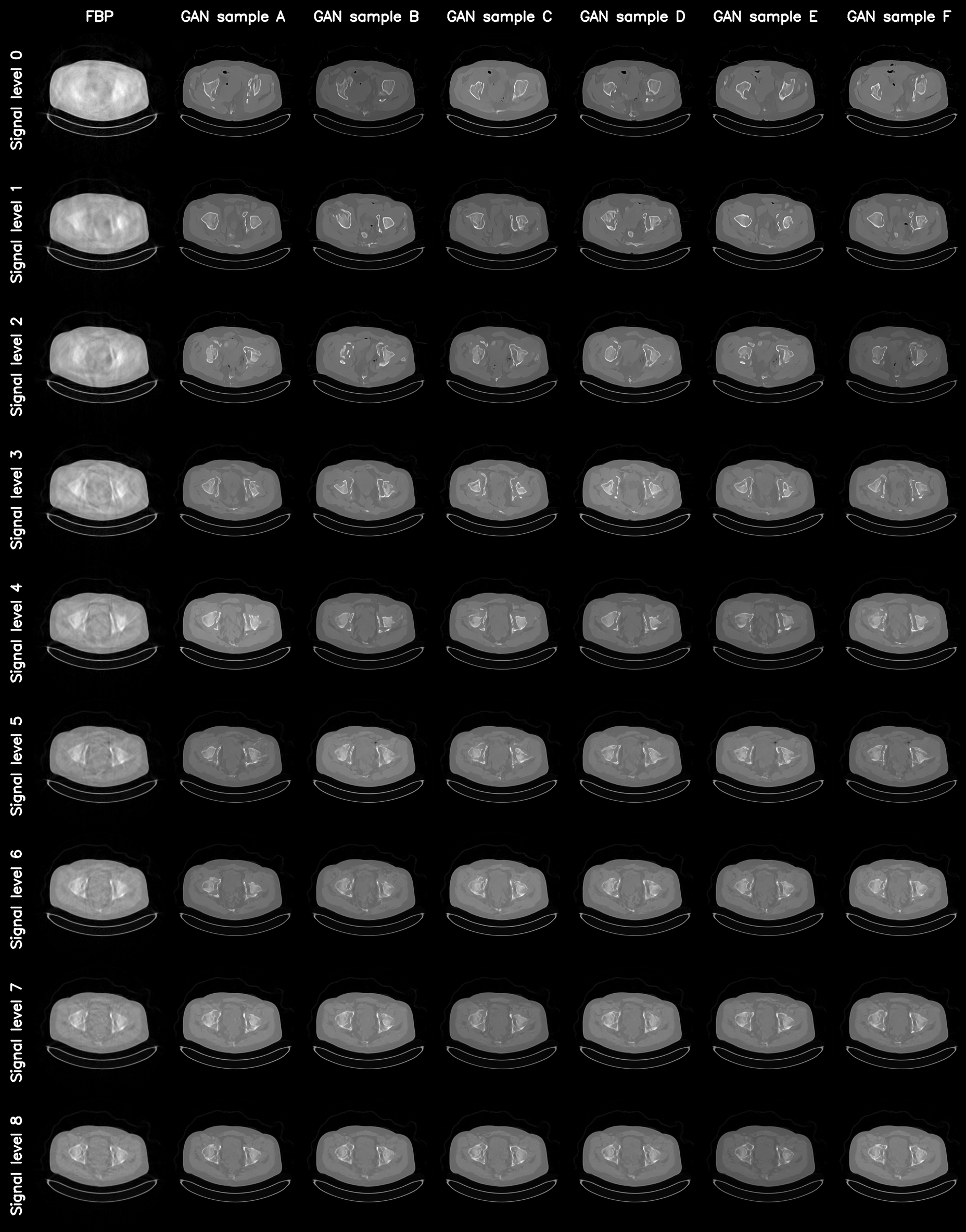}
\end{figure}

\begin{figure}
    \centering
    \includegraphics[width=\textwidth]{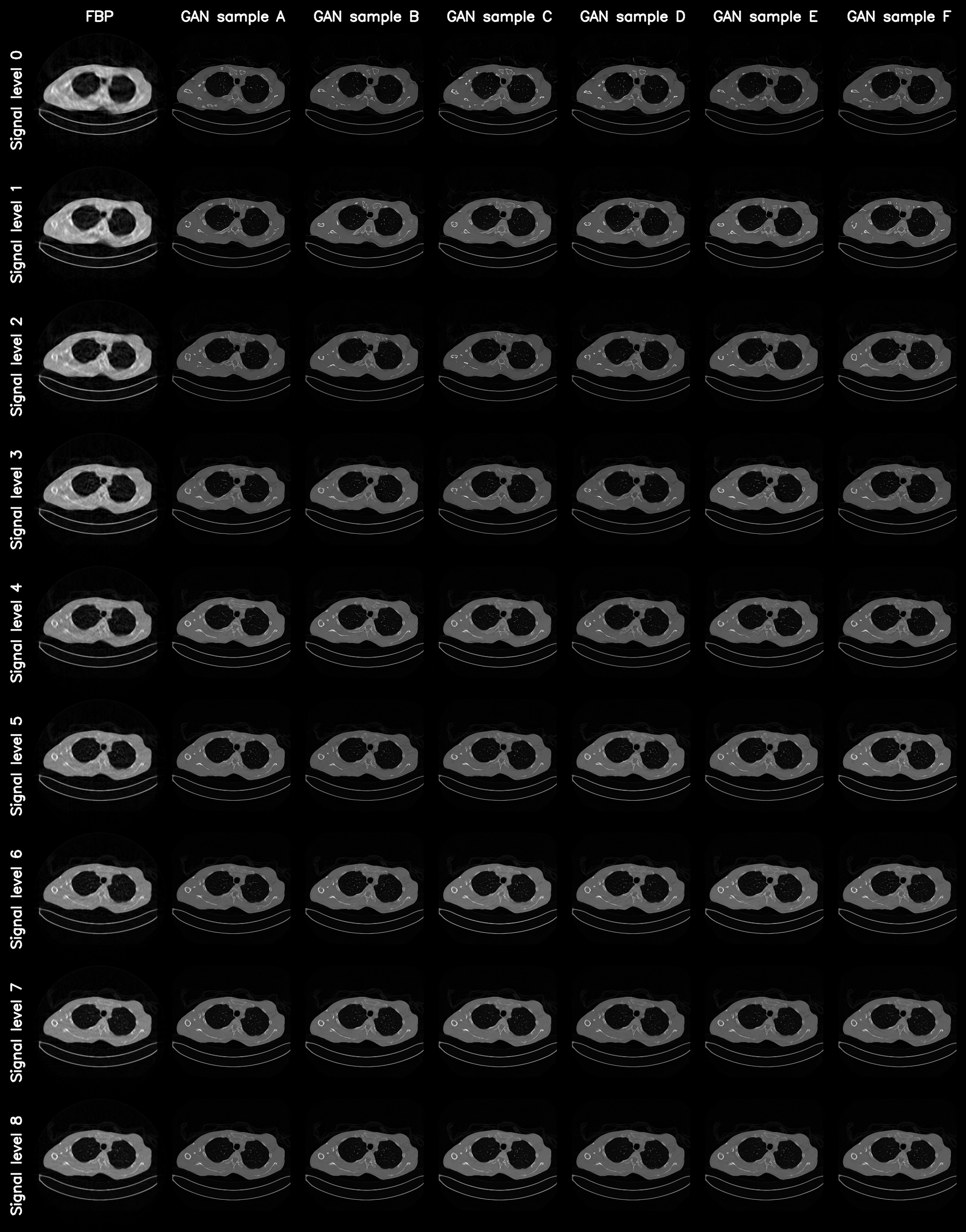}
\end{figure}

\begin{figure}
    \centering
    \includegraphics[width=\textwidth]{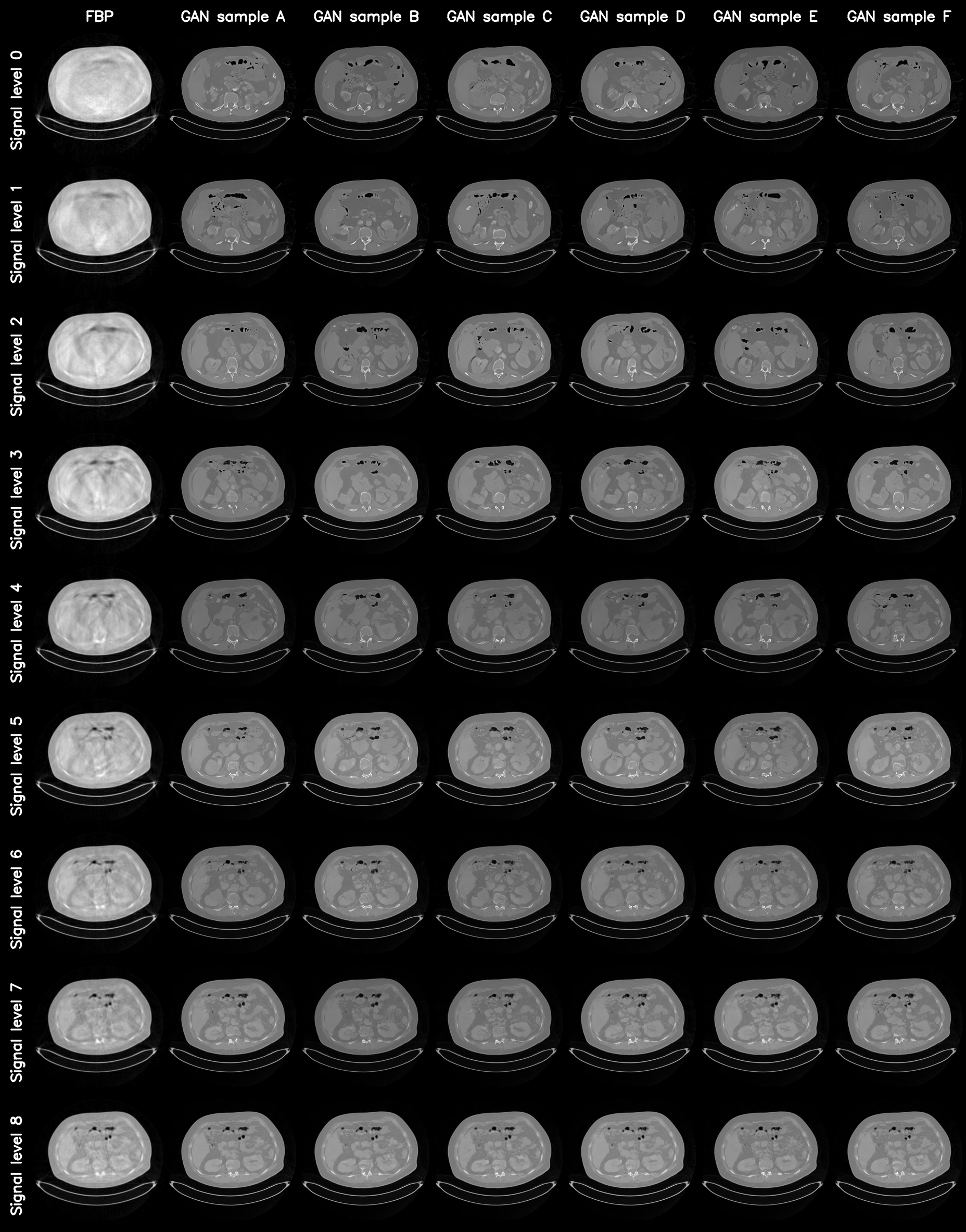}
\end{figure}

\end{document}